\documentclass[aps,pra,twocolumn,amsmath,amssymb,nofootinbib,superscriptaddress]{revtex4}

\newcommand{\bra}[1]{\langle#1|}
\newcommand{\ket}[1]{|#1\rangle}

\usepackage[dvips]{graphicx}
\usepackage{mathrsfs}

\begin{document}

\title{Practical effects in cat state breeding}

\author{Peter P. Rohde}
\email[]{rohde@physics.uq.edu.au}
\homepage{http://www.physics.uq.edu.au/people/rohde/}
\affiliation{Centre for Quantum Computer Technology, Department of Physics\\ University of Queensland, Brisbane, QLD 4072, Australia}

\author{Austin P. Lund}
\affiliation{Centre for Quantum Computer Technology, Department of Physics\\ University of Queensland, Brisbane, QLD 4072, Australia}

\date{\today}

\frenchspacing

\begin{abstract}
A cat state is a superposition of macroscopically distinct states. In quantum optics one such type of state is a superposition of distinct coherent states. Recently, a protocol has been proposed for preparing large optical cat states from a resource of smaller  ones. We consider the effects of mode-mismatch and loss in the preparation of large cat states using this protocol with a view to understand experimental limitations. (This paper is written in an experimental rapid communication format).
\end{abstract}


\maketitle

A cat state is defined as a superposition of two distinct macroscopic states. In optics\footnote{Optical cat states have uses in quantum information processing \cite{bib:Gilchrist04, bib:JeongRalph05}. The preparation of small optical cat states has been demonstrated in \cite{bib:Ourjoumtsev06}.} we use distinct coherent states\footnote{For large $\alpha$, $\langle\alpha|-\alpha\rangle\to 0$. Thus, for large $\alpha$ the states are macroscopically distinct.},
\begin{equation} \label{eq:cat_state}
\ket{cat} = \ket{\alpha} \pm \ket{-\alpha},
\end{equation}
where $\pm$ determines the parity\footnote{An odd (even) cat state is one with only odd (even) photon number terms in their expansion. These are defined as the negative and positive superpositions respectively.}.

\begin{figure}[!htb]
\includegraphics[width=0.5\columnwidth]{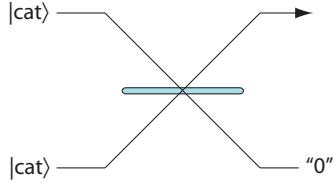}
\caption{Breeding a larger cat state from two smaller ones. One of the small cats is even parity, while the other is odd parity. They are mixed on a 50/50 beamsplitter and one of the outputs is conditioned upon detecting no photons.} \label{fig:breeder}
\end{figure}

Lund \emph{et al.} \cite{bib:Lund04} demonstrated that larger cat states can be prepared from two smaller ones using a `breeding' protocol. A slight variation\footnote{In the original protocol the `0' conditioning is not performed directly, but using an interference effect. We model the conditioning directly for simplicity. We expect our later results for state fidelity to represent an upper bound on the achievable fidelity in the original scheme, since there there is an additional opportunity for mode-mismatch to arise.} on their protocol is shown in Fig. \ref{fig:breeder}.
We consider the effects of mode-mismatch and loss on this protocol.

\section{Mode-mismatch}

First we model the temporal/spectral structure of photons using using mode-operators \cite{bib:RohdeMauererSilberhorn06}, which create photons according to a particular spectral wave-function $\psi(\omega)$,
\begin{equation}
\ket{1}_\psi = \int \psi(\omega) \hat{a}^\dag(\omega) \, \mathrm{d}\omega \, \ket{0} = \hat{A}^\dag_\psi\ket{0}.
\end{equation}

Let the state incident at one input be characterized by mode-function $\psi(\omega)$, and at the other input by $\phi(\omega)$. The input consists of one even cat state and one odd cat state, in distinct spectral modes,
\begin{equation}
\ket\psi_\mathrm{in} = (\ket\alpha_\psi - \ket{-\alpha}_\psi)\otimes(\ket\alpha_\phi + \ket{-\alpha}_\phi)
\end{equation}

We apply the beamsplitter transformation to this state and perform the conditioning. Finally, we trace out the spectral modes orthogonal to the desired one, $\psi$, to obtain the effective output state, i.e. the state that will be `seen' by homodyne tomography (see Appendix \ref{sec:propagation_proof}).

\begin{figure}[!htb]
\includegraphics[width=0.6\columnwidth]{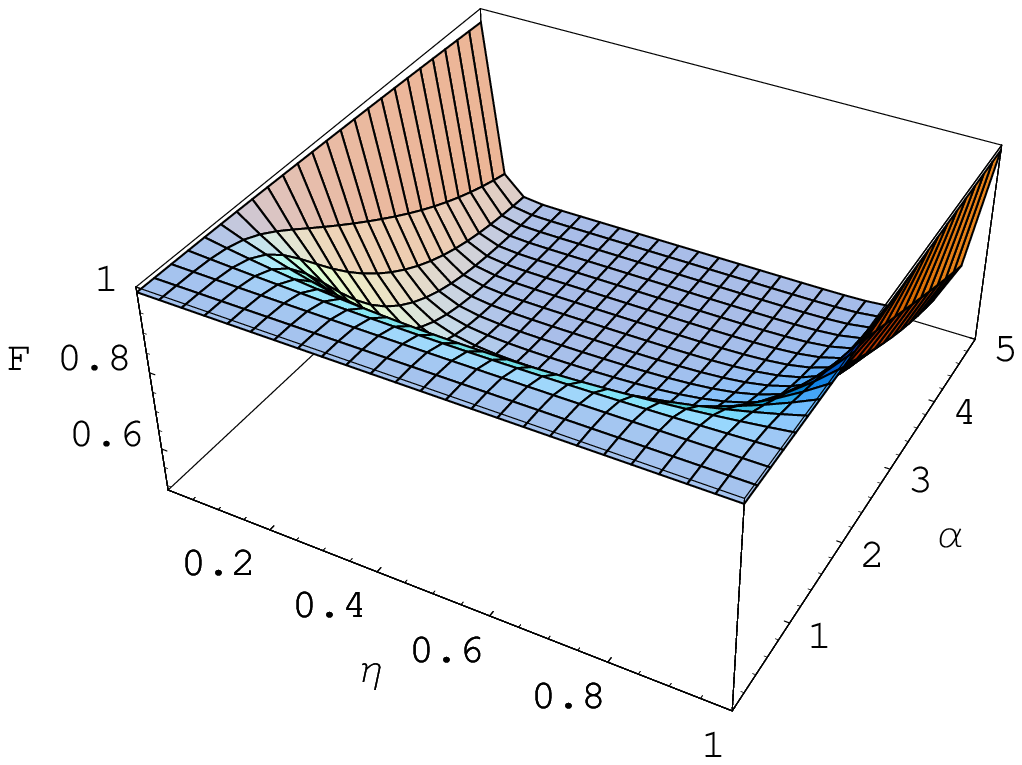}
\includegraphics[width=0.6\columnwidth]{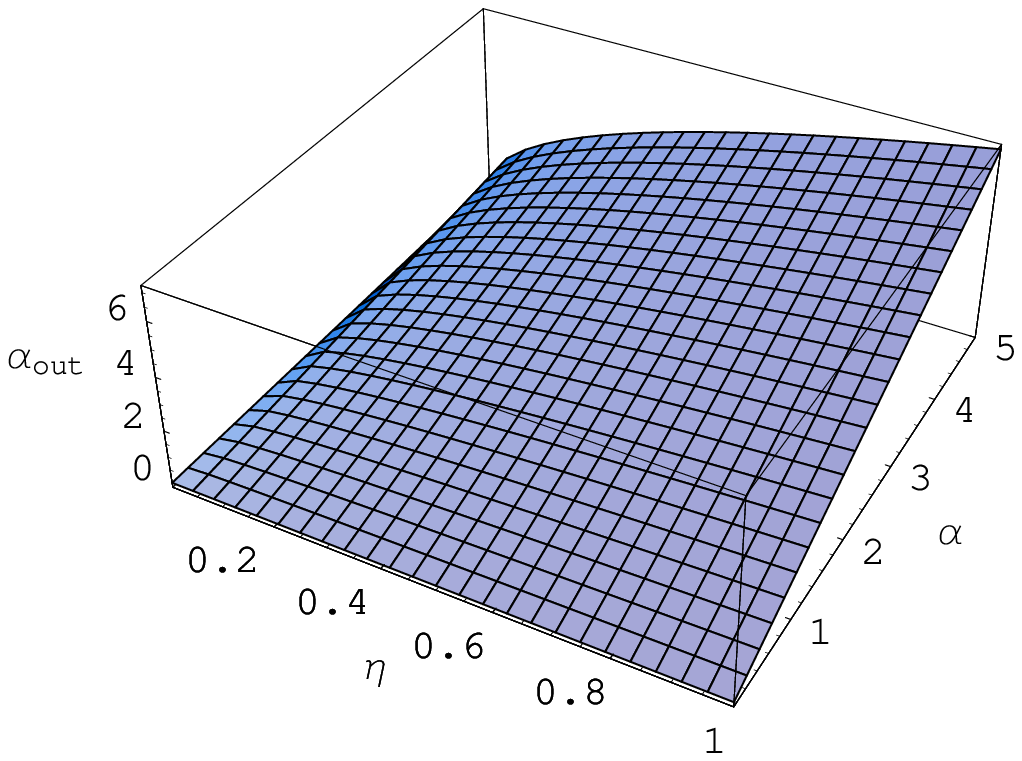}
\caption{(top) Fidelity of the effective output state. (bottom) Magnitude of the prepared cat state. $\eta$ is the mode-overlap at the beamsplitter. $\alpha$ is the magnitude of the incident states.} \label{fig:fidelity_and_alpha}
\end{figure}
In Fig. \ref{fig:fidelity_and_alpha} we plot the fidelity of the effective state compared to the expected state, and its magnitude. As $\alpha$ increases, so does the rate at which fidelity degrades as a function of mode-overlap. Thus, the scheme becomes infeasible for large cat magnitudes.

\begin{figure}[!htb]
\includegraphics[width=0.7\columnwidth]{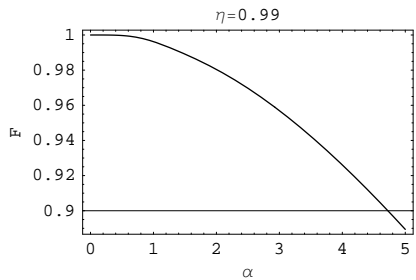}
\caption{Fidelity of the effective output state with mode-overlap fixed at $\eta=0.99$.} \label{fig:cross_section}
\end{figure}

A cross section of the fidelity plot is shown in Fig. \ref{fig:cross_section}, with mode-overlap fixed at $\eta=0.99$, which is roughly in the regime of present-day experiments.
Beyond $\alpha\approx 1$ the fidelity drops rapidly with the magnitude of the incident state. Roughly speaking, to obtain a fidelity of 90\% with a mode-overlap of 99\%, we are limited to incident states of magnitude $\alpha \leq 4.7$.

Note that our calculations assume a resource of \mbox{$F=1$} cat states. Thus, in practise we expect our plots to represent and upper bound on achievable fidelities.

\section{Loss}
Next consider the effects of loss on a prepared cat state. This was first considered in Ref. \cite{bib:Glancy04}. We model loss as a beamsplitter acting on a state of the form shown in Eq. \ref{eq:cat_state}.

Applying a beamsplitter with reflectivity $\eta$ and tracing out the reflected mode we obtain
\begin{eqnarray}
\hat\rho_\mathrm{loss} &=& \ket{\sqrt{\eta^\prime}\alpha}\bra{\sqrt{\eta^\prime}\alpha} + \ket{-\sqrt{\eta^\prime}\alpha}\bra{-\sqrt{\eta^\prime}\alpha}\nonumber\\
&+& \gamma \ket{\sqrt{\eta^\prime}\alpha}\bra{-\sqrt{\eta^\prime}\alpha} + \gamma \ket{-\sqrt{\eta^\prime}\alpha}\bra{\sqrt{\eta^\prime}\alpha},
\end{eqnarray}
where $\eta^\prime = 1-\eta$, and $\gamma= \langle\sqrt{\eta}\alpha|-\sqrt{\eta}\alpha\rangle=e^{-2\eta\alpha^2}$. This is a superposition of the coherent states $\ket{\eta^\prime\alpha}$ and $\ket{-\eta^\prime\alpha}$, where the coherence between the two terms is determined\footnote{When $\gamma=1$ we have a perfect cat state, whereas when $\gamma=0$ we have a mixture of the two terms.} by $\gamma$.

We calculate the fidelity between $\hat\rho_\mathrm{loss}$ and the cat state with magnitude\footnote{This choice of magnitude maximizes the fidelity.} $\sqrt{\eta^\prime}\alpha$. It can be shown that the fidelity is then
\begin{equation}
F = \frac{1}{2}\left(1+e^{-2\alpha^2\eta}\right).
\end{equation}

\begin{figure}[!htb]
\includegraphics[width=0.6\columnwidth]{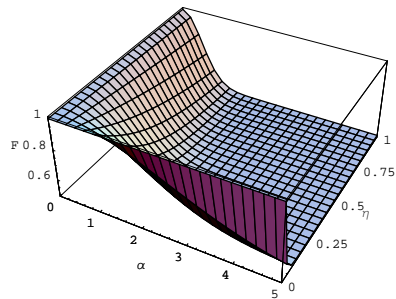}
\caption{Fidelity against loss and cat state magnitude.} \label{fig:loss_eta_alpha}
\end{figure}

\begin{figure}[!htb]
\includegraphics[width=0.6\columnwidth]{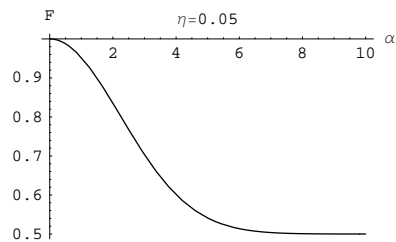}
\caption{Fidelity against cat state magnitude with loss rate fixed at 5\%.} \label{fig:loss_alpha}
\end{figure}

The fidelity is plotted if Figs. \ref{fig:loss_eta_alpha} and \ref{fig:loss_alpha}. The rate at which fidelity drops with loss increases dramatically with magnitude. Roughly speaking, with a loss rate of 5\%, if we wish to maintain the fidelity of a cat state at at least 90\%, the magnitude can be at most $\alpha\approx 1.5$.

\begin{acknowledgments}
We thank Timothy Ralph and Alex Lvovsky for helpful discussions. This work was supported by the Australian Research Council and Queensland State Government. We acknowledge partial support by the DTO-funded U.S. Army Research Office Contract No. W911NF-05-0397.
\end{acknowledgments}

\appendix

\section{Proof of output state with mode-mismatch} \label{sec:propagation_proof}
The incident cat states are of the form shown in Eq. \ref{eq:cat_state} (up to normalisation). We begin with one odd parity and one even parity state. The second copy undergoes mode-mismatch, following which it resides in the mode characterized by $\phi$.

We decompose the state into components in mode $\psi$ and orthogonal to mode $\psi$, which we label $\bar\psi$. When the states interfere on the beamsplitter, only the overlapping components of the wave-function will interact \cite{bib:RohdeMauererSilberhorn06}. Thus, we decompose the second wave-function into overlapping and non-overlapping components,
\begin{equation}
\phi(\omega) = \sqrt{\eta}\psi(\omega) + \sqrt{1-\eta}\bar\psi(\omega),
\end{equation}
where $\eta$ measures the mode-overlap. Since the modes $\psi(\omega)$ and $\bar\psi(\omega)$ are orthogonal, this effectively implements a beamsplitter transformation. For a coherent state this is
\begin{equation} \label{eq:BS_trans}
\ket\alpha_\phi \to \ket{\sqrt{\eta}\alpha}_\psi\ket{\sqrt{1-\eta}\alpha}_{\bar\psi}
\end{equation}
Thus, the second copy of the input state has the form
\begin{equation}
\ket{\alpha^\prime,\beta^\prime} \pm \ket{-\alpha^\prime, -\beta^\prime}
\end{equation}
where $\alpha^\prime = \sqrt{\eta} \alpha$ and $\beta^\prime = \sqrt{1-\eta} \alpha$. We have used the shorthand $\ket{\alpha,\beta} = \ket{\alpha}_\psi\ket{\beta}_{\bar\psi}$. Thus, the input state to the cat booster device is
\begin{equation}
\ket{\psi_\mathrm{in}} = \left(\ket{\alpha,0} - \ket{-\alpha,0}\right) \otimes
\left(\ket{\alpha^\prime,\beta^\prime}+\ket{-\alpha^\prime,-\beta^\prime}\right).
\end{equation}
To simplify the analysis, we assume that the experimenter has created matched coherent states which maximize interference. That is, $\alpha_A = \alpha_B^\prime$, where $A$ and $B$ denote the two spatial modes. This can be done by adjusting the strength of the coherent amplitude of the cat states which are input into the device.

This state is mixed on a 50/50 asymmetric beamsplitter. The output state is
\begin{eqnarray}
\ket{\psi_\mathrm{out}} &=& \ket{\sqrt{2}\alpha,0,\beta^\prime/\sqrt{2},\beta^\prime/\sqrt{2}} \nonumber\\
&-& \ket{0,-\sqrt{2}\alpha,\beta^\prime/\sqrt{2},\beta^\prime/\sqrt{2}} \nonumber\\
&+& \ket{0,\sqrt{2}\alpha,-\beta^\prime/\sqrt{2},-\beta^\prime/\sqrt{2}}\nonumber\\
&-& \ket{-\sqrt{2}\alpha,0,-\beta^\prime/\sqrt{2},-\beta^\prime/\sqrt{2}}.
\end{eqnarray}
where $\ket{\alpha,\beta,\gamma,\delta}= \ket{\alpha}_{A,\psi} \ket{\beta}_{B,\psi} \ket{\gamma}_{A,\bar\psi} \ket{\delta}_{B,\bar\psi}$.

Detection of the output mode requires that zero photons are measured in mode $B$. Two components of this state contribute to the detection result -- modes 2 and 4. These modes must have zero photons total for the measurement to give the required outcome. So projecting these modes onto the zero photon state gives
\begin{eqnarray} \label{state_after_measurement}
\ket{\psi_\mathrm{proj}}&=&\ket{\sqrt{2}\alpha,\beta^\prime/\sqrt{2}} - e^{-\alpha^2} \ket{0,\beta^\prime/\sqrt{2}} \nonumber\\
&+&e^{-\alpha^2} \ket{0,-\beta^\prime/\sqrt{2}} - \ket{-\sqrt{2}\alpha,-\beta^\prime/\sqrt{2}},\nonumber\\
\end{eqnarray}
where $\ket{\alpha,\beta} = \ket{\alpha}_{A,\psi} \ket{\beta}_{A,\bar\psi}$.

For later convienence, we rewrite this state as
\begin{equation} \label{state_after_measurement_simple}
\ket{\psi_\mathrm{proj}} = \ket{\Psi_+} \ket{\beta^\prime/\sqrt{2}} - \ket{\Psi_-} \ket{-\beta^\prime/\sqrt{2}},
\end{equation}
where $\ket{\Psi_\pm} = \ket{\pm \sqrt{2}\alpha} - e^{-\alpha^2} \ket{0}$.
Here the second mode is that which is not matched to the analyser.

When a state like this is measured using, for example, homodyne tomography, the measurement device will be tuned so as to maximize mode-overlap with the state. This is optimized by tuning to the mode defined by $\psi$, the desired mode. A homodyne measurement will be insensitive to the component of the state orthogonal to the chosen mode, in which case the information from that mode is lost. Thus, we calculate the state seen by the measurement device by tracing out mode $\bar\psi$. Thus, the effective output state is
\begin{eqnarray}
\hat\rho_\mathrm{eff} &=& \mathrm{tr}_{\bar\psi}(\hat\rho_\mathrm{proj}) \nonumber\\
&=& \ket{\Psi_+}\bra{\Psi_+} + \ket{\Psi_-}\bra{\Psi_-} \nonumber\\
&-& \langle\beta^\prime/\sqrt{2} | -\beta^\prime/\sqrt{2}\rangle \left(\ket{\Psi_+}\bra{\Psi_-} + \ket{\Psi_-}\bra{\Psi_+} \right).\nonumber\\
\end{eqnarray}
Expressed in terms of coherent states,
\begin{eqnarray}
\hat\rho_\mathrm{eff} &=& \ket{\sqrt{2}\alpha}\bra{\sqrt{2}\alpha}
+ \ket{-\sqrt{2}\alpha}\bra{-\sqrt{2}\alpha} \nonumber\\
&-& e^{-\beta^2}(\ket{\sqrt{2}\alpha}\bra{-\sqrt{2}\alpha}
- \ket{-\sqrt{2}\alpha}\bra{\sqrt{2}\alpha}) \nonumber\\
&+& e^{-\alpha^2}(e^{-\beta^2}-1)(\ket{\sqrt{2}\alpha}\bra{0}
+ \ket{0}\bra{\sqrt{2}\alpha} \nonumber\\
&+& \ket{-\sqrt{2}\alpha}\bra{0}
+ \ket{0}\bra{-\sqrt{2}\alpha})\nonumber\\
&+& 2e^{-2\alpha^2}(1-e^{-\beta^2})\ket{0}\bra{0}.
\end{eqnarray}
Note that the output cat state has been mixed with the vacuum state. The non-zero overlap of the coherent states with the vacuum state has also introduced coherences between the cat state and the vacuum. Note also that as $\beta \rightarrow 0$ the state reduces to
\begin{equation}
\hat\rho_\mathrm{eff} = (\ket{\sqrt{2}\alpha} + \ket{-\sqrt{2}\alpha}) \otimes H.c.,
\end{equation}
which is the odd cat state as desired.

\bibliography{paper}

\begin{thebibliography}{6}
\expandafter\ifx\csname natexlab\endcsname\relax\def\natexlab#1{#1}\fi
\expandafter\ifx\csname bibnamefont\endcsname\relax
  \def\bibnamefont#1{#1}\fi
\expandafter\ifx\csname bibfnamefont\endcsname\relax
  \def\bibfnamefont#1{#1}\fi
\expandafter\ifx\csname citenamefont\endcsname\relax
  \def\citenamefont#1{#1}\fi
\expandafter\ifx\csname url\endcsname\relax
  \def\url#1{\texttt{#1}}\fi
\expandafter\ifx\csname urlprefix\endcsname\relax\def\urlprefix{URL }\fi
\providecommand{\bibinfo}[2]{#2}
\providecommand{\eprint}[2][]{\url{#2}}

\bibitem[{\citenamefont{Gilchrist et~al.}(2004)\citenamefont{Gilchrist, Nemoto,
  Munro, Ralph, Glancy, Braunstein, and Milburn}}]{bib:Gilchrist04}
\bibinfo{author}{\bibfnamefont{A.}~\bibnamefont{Gilchrist}},
  \bibinfo{author}{\bibfnamefont{K.}~\bibnamefont{Nemoto}},
  \bibinfo{author}{\bibfnamefont{W.~J.} \bibnamefont{Munro}},
  \bibinfo{author}{\bibfnamefont{T.~C.} \bibnamefont{Ralph}},
  \bibinfo{author}{\bibfnamefont{S.}~\bibnamefont{Glancy}},
  \bibinfo{author}{\bibfnamefont{S.~L.} \bibnamefont{Braunstein}},
  \bibnamefont{and} \bibinfo{author}{\bibfnamefont{G.~J.}
  \bibnamefont{Milburn}}, \bibinfo{journal}{J. Opt. B}
  \textbf{\bibinfo{volume}{6}}, \bibinfo{pages}{S828} (\bibinfo{year}{2004}).

\bibitem[{\citenamefont{Jeong and Ralph}(2005)}]{bib:JeongRalph05}
\bibinfo{author}{\bibfnamefont{H.}~\bibnamefont{Jeong}} \bibnamefont{and}
  \bibinfo{author}{\bibfnamefont{T.~C.} \bibnamefont{Ralph}}
  (\bibinfo{year}{2005}), \eprint{quant-ph/0509137}.

\bibitem[{\citenamefont{Ourjoumtsev et~al.}(2006)\citenamefont{Ourjoumtsev,
  Tualle-Brouri, Laurat, and Grangier}}]{bib:Ourjoumtsev06}
\bibinfo{author}{\bibfnamefont{A.}~\bibnamefont{Ourjoumtsev}},
  \bibinfo{author}{\bibfnamefont{R.}~\bibnamefont{Tualle-Brouri}},
  \bibinfo{author}{\bibfnamefont{J.}~\bibnamefont{Laurat}}, \bibnamefont{and}
  \bibinfo{author}{\bibfnamefont{P.}~\bibnamefont{Grangier}},
  \bibinfo{journal}{Science} \textbf{\bibinfo{volume}{312}},
  \bibinfo{pages}{83} (\bibinfo{year}{2006}).

\bibitem[{\citenamefont{Lund et~al.}(2004)\citenamefont{Lund, Jeong, Ralph, and
  Kim}}]{bib:Lund04}
\bibinfo{author}{\bibfnamefont{A.~P.} \bibnamefont{Lund}},
  \bibinfo{author}{\bibfnamefont{H.}~\bibnamefont{Jeong}},
  \bibinfo{author}{\bibfnamefont{T.~C.} \bibnamefont{Ralph}}, \bibnamefont{and}
  \bibinfo{author}{\bibfnamefont{M.~S.} \bibnamefont{Kim}},
  \bibinfo{journal}{Phys. Rev. A} \textbf{\bibinfo{volume}{70}},
  \bibinfo{pages}{020101(R)} (\bibinfo{year}{2004}).

\bibitem[{\citenamefont{Rohde et~al.}(2006)\citenamefont{Rohde, Mauerer, and
  Silberhorn}}]{bib:RohdeMauererSilberhorn06}
\bibinfo{author}{\bibfnamefont{P.~P.} \bibnamefont{Rohde}},
  \bibinfo{author}{\bibfnamefont{W.}~\bibnamefont{Mauerer}}, \bibnamefont{and}
  \bibinfo{author}{\bibfnamefont{C.}~\bibnamefont{Silberhorn}}
  (\bibinfo{year}{2006}), \eprint{quant-ph/0609004}.

\bibitem[{\citenamefont{Vasconcelos and Ralph}(2004)}]{bib:Glancy04}
\bibinfo{author}{\bibfnamefont{S.~G. H.~M.} \bibnamefont{Vasconcelos}}
  \bibnamefont{and} \bibinfo{author}{\bibfnamefont{T.~C.} \bibnamefont{Ralph}},
  \bibinfo{journal}{Phys. Rev. A}  (\bibinfo{year}{2004}).

\end{thebibliography}

\end{document}